\documentclass[final,1p,times]{elsarticle}

 \usepackage{graphicx}

\usepackage{amssymb}

\journal{Nuclear Physics A}

\begin{document}

\begin{frontmatter}


 

\title{Feshbach Resonance due to Coherent $\Lambda$-$\Sigma$ Coupling in $_{\Lambda}^{7}$He}

\author[label1]{San San Mon}
\author[label1]{Tin Tin Nwe}
\author[label2]{Khin Swe Myint}
\ead{pro-rector@mptmail.net.mm}
\author[label3]{Y. Akaishi}
\address[label1]{Department of Physics, Mandalay University, Myanmar.}
\address[label2]{Pro-Rector, Mandalay University, Myanmar.}
\address[label3]{College of Science and Technology, Nihon University, Chiba, Japan and RIKEN Nishina Center, Saitama, Japan.}
\begin{abstract}
Coherent $\Lambda$-$\Sigma$ coupling effect in $_{\Lambda}^{7}$He is analyzed within three-body framework of two coupled channels, $\Lambda$-$t$-$t$ and $\Sigma$-$\tau$-$t$, where $\tau$ represents trinulceon which is either $^{3}$H or $^{3}$He. The hyperon-trinucleon ($Y\tau$) and trinucleon-trinucleon ($\tau\tau$) interactions are derived by folding $G$-matrices of $YN$ and $NN$ interactions with trinucleon density distributions. It is found that the binding energy of $_{\Lambda}^{7}$He is 4.04 MeV below the $\Lambda$+$t$+$t$ threshold without $\Lambda$-$\Sigma$ coupling and the binding energy is increased to 4.46 MeV when the coupling effect is included. This state is 7.85 MeV above the $^{6}$He+$\Lambda$ threshold and it may have a chance to be observed as a Feshbach resonance in $^{7}$Li~$(e,e'K^{+})$ $_{\Lambda}^{7}$He experiment done at Jefferson Lab. 
\end{abstract}

\begin{keyword}

Feshbach resonance: coherent $\Lambda$-$\Sigma$ coupling: hyperon-trinucleon interaction 

\PACS
\end{keyword}
\end{frontmatter}

\section{Introduction}

Significance of $\Lambda$-$\Sigma$ coupling effect in binding mechanism of light $\Lambda$-hypernuclei has long been recognized and discussed in the references \cite{YN,BF}. Admixture of $\Sigma$ states in $\Lambda$-hypernuclei is probably an important aspect of hypernuclear dynamics. There are two coupling schemes namely incoherent and coherent $\Lambda$-$\Sigma$ couplings \cite{YA}. Incoherent $\Lambda$-$\Sigma$ coupling means a nucleon changes to an excited level after the interaction, while the other process where a nucleon remains in its ground state after converting $\Lambda$ to $\Sigma$, is called coherent $\Lambda$-$\Sigma$ coupling. In the latter case, all the nucleons have an equal chance to interact with the converted $\Sigma$ and coupling effect contributed from each nucleon is added coherently. Harada \cite{TH} has successfully fitted the experimental spectra of $^{4}$He (stopped $K^{-}, \pi^{-}$) \cite{RH} and $^{4}$He (in-flight $K^{-},\pi^{-})$ \cite{TN} production reactions by taking into account the coherent $\Lambda$-$\Sigma$ coupling effect. Furthermore, all the s-shell $\Lambda$ hypernuclear binding energies are well reproduced only after the coherent $\Lambda$-$\Sigma$ coupling effect has been included \cite{YA,HN}. It has been found that the coherent coupling contribution is significantly large on the order of 1 MeV in $^{4}_{\Lambda}$H and $^{4}_{\Lambda}$He ground states. 

\section{Coupled-channel three-body cluster model of $^{7}_{\Lambda}$He}

Having considered the above mentioned findings, we analyze a structure of $^{7}_{\Lambda}$He in continuum by using three-body model of $\Lambda$-$t$-$t$, $\Sigma^0$-$t$-$t$ and $\Sigma^-$-$h$-$t$ coupled channels to investigate the coherent $\Lambda$-$\Sigma$ coupling effect. The coupling between $\Lambda t$-$\Sigma^{0}t$ gives coherent $\Lambda$-$\Sigma$ coupling, while Lane term of $\Sigma^{0}t$-$\Sigma^{-}h$ coupling plays a significant role in forming $^{4}_{\Sigma}$H \cite{THa}. All these couplings are included in our analysis. To solve three-body calculation, we employ Kamimura's coupled rearrangement-channel method \cite{MK}. 

Three-body Hamiltonian of the $\Lambda$-$t$-$t$ diagonal part, which we explicitly show for explanation here, is 
\begin{equation}
H_{\Lambda tt} = -\frac{\hbar^{2}}{2M_{c}}\Delta_{\vec{R}_{c}}- \frac{\hbar^{2}}{2\mu_{c}}\Delta_{\vec{r}_{c}}+\{V_{tt}(\vec{r}_{1})+ V_{\Lambda t}(\vec{r}_{2})+ V_{t\Lambda}(\vec{r}_{3})\}+ V_{\mbox{Pauli}}(\vec{r}_{1},\vec{r'}_{1}),
\end{equation}
where $V_{\mbox{Pauli}}$ expresses Pauli exclusion effect between two tritons. In orthogonality-condition model (OCM) \cite{SS},
\begin{equation}
V_{\mbox{Pauli}} (\vec{r},\vec{r'}) =\lim_{\lambda \to \infty}\lambda \sum_{\mbox{f}}|\,\,\Phi_{\mbox{f}}(\vec{r})\,\,\rangle\langle\,\Phi_{\mbox{f}}\,\,(\vec{r'})\,| ,
\end{equation}
where $\Phi_{\mbox{f}}(\vec{r})$ is the Pauli forbidden state. Total wave function of the $\Lambda$-$t$-$t$ channel is expanded in Gaussian bases which are spanned over three rearrangement-channels as follows,
\begin{equation}
\Psi_{\Lambda tt}(\vec{r},\,\vec{R})= \sum^{3}_{c=1}\,\sum_{i_{c}\,j{c}}\,A^{(c)}_{i_{c}\,j_{c}}\,e^{-(\frac{\vec{r}_{c}}{b_{i}})^2}\, e^{-(\frac{\vec{R}_{c}}{b_{i}})^2}\ .
\end{equation}
Wave functions of the other channels $\Sigma^0$-$t$-$t$ and $\Sigma^-$-$h$-$t$ are treated in a similar way.
\begin{table}[b]
\begin{center}
\renewcommand{\arraystretch}{1.2}
\begin{tabular}{|c|c|c|c|c|c|c|}
	\hline
State & \multicolumn{3}{|c|}{$^{4}_{\Lambda}$H(0$^{+})$} & \multicolumn{3}{|c|} {$^{4}_{\Lambda}$H(1$^{+})$}\cr 
\hline
$k$ & $V^{S=0}_{k}(\Lambda t$-$\Lambda t)$ & $V^{S=0}_{k}(\Sigma t$-$\Sigma t)$  & $V^{S=0}_{k}(\Lambda t$-$\Sigma t)$  & $V^{S=1}_{k}(\Lambda 	t$-$\Lambda t)$  & $V^{S=1}_{k}(\Sigma t$-$\Sigma t)$  & $V^{S=1}_{k}(\Lambda t$-$\Sigma t)$ \cr
       	\hline
      
	  1 & 1.7284 & 9.4720 & -3.5575 & 0.36869 & 1.9558 & -0.16822 \cr
	  2 & 50.838 & 69.234 & 4.2647 & 43.237 & 65.877 & 7.1105\cr
	  3 & -63.595 & -105.09 & 32.682 & -57.877 & -44.391 & 0.70109\cr
	  4 & 6.2861 & 10.130 & -4.0631 & 5.1858 & 2.1056 & -1.3503\cr
	  5 & -1.1202 & -2.3001 & 0.8537 & -0.86971 & -0.61958 & 0.12653\cr 
	  \hline
	\end{tabular}
	\end{center}
\caption{The hyperon-trinucleon interactions in MeV for $^{4}_{\Lambda}$H(0$^{+})$ and $^{4}_{\Lambda}$H(1$^{+})$. The range parameters are $\mu_{1}=1.00$ fm, $\mu_{2}=1.37$ fm, $\mu_{3}=1.87$ fm, $\mu_{4}=2.56$ fm, $\mu_{5}=3.50$ fm.}
\label{Int}
\end{table} 

The $YN$ interaction used in our computations is a phase equivalent potential of the Nijmegen model-D $YN$ potential \cite{MM}. Then, hyperon-trinucleon potentials are obtained by folding the effective interaction, i.e. $G$-matrix of the above $YN$ potential with trinucleon density distributions \cite{SA}. They are expressed in five-range Gaussian form, the range and strength parameters of which are slightly modified so as to reproduce the empirical $\Lambda$ binding energy of $^{4}_{\Lambda}$H(0$^{+})$ and $^{4}_{\Lambda}$H(1$^{+})$, and the expansion coefficients for $I=1/2$ are given in Table \ref{Int}. Trinucleon-trinucleon ($\tau$-$\tau$) interaction is obtained by doubly folding $G$-matrix of Tamagaki's OPEG $NN$ potential with trinucleon density distributions. This $\tau$-$\tau$ potential is spin-isospin dependent, and does not give any bound state of triton-triton two-body system in OCM treatment.

\section{Results and discussions}

\begin{figure}[h]
\centering
\includegraphics[width=10cm]{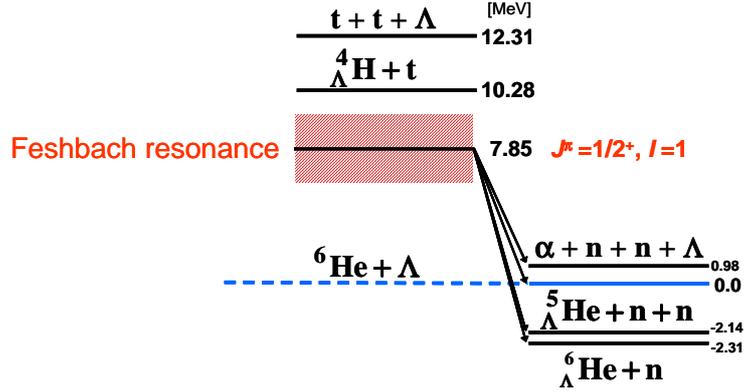}
\vspace{0cm}
\caption{\label{level} 
The obtained Feshbach-resonance state of $t+t+\Lambda$. It is shown together with the thresholds of various channels. }
\end{figure}

From our calculation, a bound state is found to be at $4.46$ MeV below the $t+t+\Lambda$ threshold and about $7.85$ MeV above the $^{6}$He$+\Lambda$ threshold as shown in Fig. \ref{level}. It is a Feshbach resonance state \cite{HF}, because it lies in continuum region of the open channels such as  $^{6}$He$+\Lambda$, $^{6}_{\Lambda}$He$+n$, $^{5}_{\Lambda}$He$+n+n$ and $\alpha +n+n+ \Lambda$ channels.

\begin{figure}[h]
\centering
\includegraphics[width=9cm]{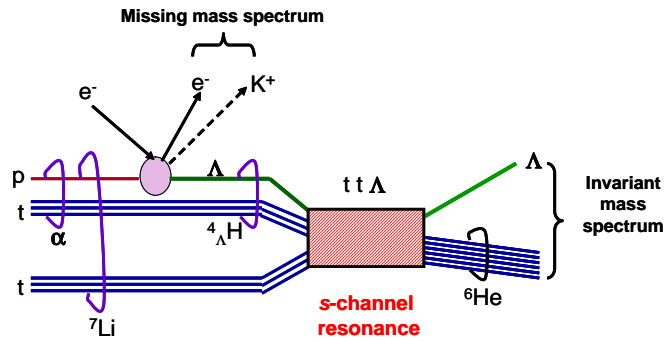}
\vspace{0cm}
\caption{\label{react} 
Production and decay mechanisms of the $tt\Lambda$ resonance state through $^7$Li$~(e,e'K^{+})$ reaction. }
\end{figure}

 A possible way to populate this resonance state, $^{7}_{\Lambda}$He$^*$, is through  $(e,e'K^{+})$ electro-production reaction on $^{7}$Li target. Formation of $^{7}_{\Lambda}$He$^*$ through the $tt\Lambda$ resonance is described with $s$-channel interaction model as shown in Fig. \ref{react}. Formation and decay spectra are analyzed, as explained in Ref. \cite{YAk}, by using Yamaguchi-type separable (i.e. $s$-channel) potential:
\begin{equation}
\langle\,\,\vec{k}\,\,|V_{ij}|\vec{k'}\,\rangle\,=\,g_{i}(\vec{k})\,U_{ij}\,g_{j}(\vec{k'}),\,\,\,\,\,\,g_{i}(\vec{k})=\frac{\Lambda^{2}_{i}}{\Lambda^{2}_{i}\,+\vec{k}^2},
\end{equation}
where $i, j = t~^4_{\Lambda}$H$, \Lambda^6$He.

Missing-mass spectrum and invariant-mass spectrum can be obtained by detecting emitted particles $e'$ and $K^{+}$ and decay particle $\Lambda$, respectively. The effect of interaction range on the missing-mass spectrum is investigated by varying the range parameter of $^{4}_{\Lambda}$H-$t$ interaction from 0.3 to 0.9 fm.

\begin{figure}[h]
\centering
\includegraphics[width=9.6cm]{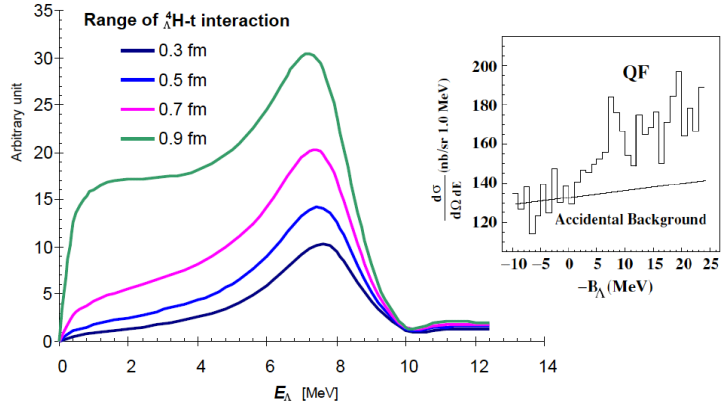}
\vspace{0cm}
\caption{\label{spect} 
Missing-mass spectrum of the $tt\Lambda$ resonance state. Experimental data is taken from Ref. \cite{LY}. }
\end{figure}

Figure \ref{spect} shows the missing-mass spectrum calculated with 3 MeV width of the $tt\Lambda$ resonance. We have compared this missing-mass spectrum with JLab experimental spectrum \cite{LY}, where a peak structure is found at about 7 MeV above the $^{6}$He+$\Lambda$ threshold, which might correspond to our resonance state. A crude explanation of why a narrow peak appears in continuum region is such that; similarity in structures between $\alpha$-$t$ and $t$-$t$ may give a strong population of $tt\Lambda$ state, while different structures between $t$-$t$ and $^{6}$He ensure the formation of quasi-stable Feshbach resonance. However, a recent experimental spectrum of $^{7}$Li~$(e,e'K^{+})~^{7}_{\Lambda}$He displays only a prominent peak below the $^{6}$He+$\Lambda$ threshold in bound region \cite{OH}. In order to clarify the possible existence of Feshbach resonance in $^{7}_{\Lambda}$He system, electro-production or equivalent experiments on $^{7}$Li target with high statistics are highly awaited.\\

Two of the authors, San San Mon and Khin Swe Myint, would like to thank the organizing committee for the support to attend the Conference.

\end{document}